# Measuring Changes in Instructor Class Design and Student Learning After the Release of Large Language Models (LLMs)

Amanda Potasznik, University of Massachusetts, Boston, USA
Daniel Haehn, University of Massachusetts, Boston, USA

***Abstract:*** Student use of Generative AI (GenAI) products in completing their classwork, with or without their professors' knowledge and/or approval, has resulted in substantial shifts in higher education. While GenAI use is widespread, its impact on student study methods, faculty course development, grade reporting, and overall learning is not well documented. This is a mixed-methods, multi-course study using retrospective quantitative analysis, instructor surveys, and anonymous student surveys at a university in the New England region of the United States. This research seeks to identify and document patterns in student and faculty perceptions of, and experiences in, the use of LLMs as a learning tool inside and outside of the university classroom. Alongside quantitative and thematic analysis of both faculty and student survey responses, historical grade data as reported to the university registrar is used to triangulate the phenomenon of learning achievement in pre- and post-LLM eras. It is hoped that this research can serve as a pilot study for a broader set of institutions. Results from this study can inform GenAI policy for professors, universities, and other educational institutions that are trying to maximize student learning in the age of AI.

## Introduction

The public release of ChatGPT in November 2022, and subsequent public releases of Copilot, Gemini, Claude, and other models marked a major increase in students' access to generative AI. Instructors across education levels and disciplines have reported changes in student work quality, assignment performance, conceptual understanding, and learning behaviors. These reports, however, are largely anecdotal, and empirical study of potential trends is difficult since traditional metrics (such as grades) are compromised: for several reasons, grades are becoming even less accurate as a proxy for underlying learning (admittedly, controversy about that particular topic has been documented in the literature for decades (Allen 2005; Cockrell 2025; Lambating and Allen 2002; Martin Sanz et al. 2017). First, there is the "cat and mouse" phenomenon (Young 2001): Instructors may redesign post-2022 assignments to be less LLM-answerable, so comparability across semesters is compromised. Some instructors have completely, if reluctantly, overhauled their courses for this purpose; others have not addressed it. Second, students who use LLMs undetected may enjoy artificially inflated grades, as the LLM-generated answers are not always identified as such by instructors and their quality may be high enough to "earn" good grades. Essentially, grades don't reflect learning because grades are high and learning is low. Finally, students whose LLM use *is* discovered may fail assignments, which means that failing grades may be assigned even when students have learned substantial amounts. Essentially, grades don't reflect learning because grades are low and learning is high.

Given these issues, universities need evidence-based assessments of learning trends over time using triangulated grade and non-grade measures. This research seeks to inform policymaking at the university level by identifying and documenting patterns in student and faculty perceptions of, and experiences in, the use of LLMs as a learning tool inside and outside of the university classroom. Alongside quantitative and thematic analysis of both faculty and student survey responses, historical grade data as reported to the university registrar is used to triangulate the phenomenon of learning achievement in pre- and post-LLM eras. The research site for this study was the University of Massachusetts, Boston; it is hoped that this research can serve as a pilot study for a broader set of institutions.

A note on verbiage: Large Language Models (LLMs) are one category of Generative AI (GenAI). "Whereas generative AI encompasses a wider scope of content generation abilities, LLMs, as a subset of generative AI, are applied to perform language-related tasks specifically. They power software that ... synthetically generates written text, such as ... helping students enhance essays, or summarizing long documents" (Sandhu 2024). Since in educational contexts, most traditional assessments incorporate written text as opposed to image generation, in this paper, the terms LLMs are GenAI are used interchangeably. Regarding timelines, we ascribe the label of "post-LLM era" to semesters after the widespread availability of GenAI products in the United States, which coincides with the public release of ChatGPT in November, 2022.

## Research Questions

The primary research question is: How have teaching and learning patterns changed in the university classroom in the post-LLM era at UMB? Sub-questions include:

1. To what extent have instructors changed their courses in light of student access to LLMs? What is the catalyst and timeline for instructors to make these changes? How do instructors perceive these changes?
2. To what extent does student LLM access manifest in student grades as reported by instructors?
3. To what extent might instructor-reported changes in assignment design in the post-LLM era align with changes in observed grade reporting patterns?
4. What self-reported learning behaviors have students adopted in the post-LLM era (e.g., reliance on AI, reduced studying, different study strategies)?
5. To what extent has students' conceptual understanding and ability to apply knowledge changed since the widespread availability of LLMs? To what extent do students underreport LLM use, and how does this affect interpretation of learning outcomes?

## Background information

In order to answer these sub-questions, we needed specific and sometimes new data sources. For sub-question one, research required new survey input because traditionally, college instructors are not required to fully document changes that they may make to their courses semester by semester. Rather, college, departmental, and university administrations have historically respected professors' expertise in their subject and, by extension, good judgement for course design. Within the guardrails of university-approved course objectives, many professors can employ their own discretion about course structure, curriculum content, and score weighting (Huang et al. 2007), resulting in substantial leeway and minimal documentation for instructors designing or updating their courses. Relatively recently, that freedom has been labeled as an "assumption" (Whittington 2024) that is not guaranteed, and is subject to ideological twists and turns of political influence ("Attacks at the State Level | AAUP," n.d.; Kamola 2024). Despite that recent development, there is little standardized, empirical evidence to describe course material and grading trends across educational institutions or even within departments that can be used to compare timeframes before and after LLM access by students.

For sub-question two, grades are documented in individual student transcripts and instructors' semesterly gradebooks, but multiple semesters' worth of historical grade data is not easily accessed; instead, reports require special permissions from and cooperation with the university registrar. For sub-question three, we need to compare the answers to sub-questions two and three. Sub-questions four and five necessitate input from students, so those questions will most likely be answered by analyzing student survey data. That analysis will also be informed by instructor survey answers. For sub-question four, we asked students via a survey to quantify and explain their perspectives on these subjects.

For sub-question five, we used data gathered from both faculty and student surveys to glean insights on their respective views about conceptual understanding and ability to apply knowledge. Admitting AI use can often bring negative consequences in academic settings, leading to social desirability bias: "systematic error in self-report measures resulting from the desire of respondents to avoid embarrassment and project a favorable image to others" (Fisher 1993, 303) To estimate rates of under-reporting of AI use for students, we took a two-pronged approach. There was direct reporting: we asked students directly to quantify their personal LLM use, and expected this phenomenon to be underreported. Then, using indirect questioning as suggested in (1993), we asked students to estimate the rates of LLM usage by their peers. The difference between these two prongs provided an underreporting index.

This research seeks to elucidate trends in both grading and student's conceptual understanding in the post-LLM era by triangulating reports of grades as submitted to the university registrar, faculty input, and student input. Further explanation of data sources and rationale are discussed in the Methods section.

## Literature Review

### Grade Reporting and Learning

Measuring student learning has long been an imperfect science. While traditional grading systems are meant to be valid measures of academic performance, they have been criticized for their perceived lack of validity (Marzano 2000; Allen 2005). Researchers have pointed out that grades are "biased" and "underestimate the true ability" of many groups of students (Walton and Spencer 2009). Still, grades continue to be the most succinct marker of student achievement for both individual assignments and overall course performance in primary, secondary, and post-secondary education institutions in the United States and abroad. Their continued usage can be attributed to several factors, from inertia to simplicity: "Grades convert what would otherwise be a very complicated type of learning assessment into a very simple and widely accepted metric, a shorthand of sorts for both professors and students that is fast, convenient, and easy to record... Grades are a widely accepted proxy for knowhow" (Schwab et al. 2018). These and other identified benefits of the traditional grading system have solidified the widespread use of grades as markers of knowledge possession and, hence, academic achievement.

Historically, at the university level, cumulative grade point averages (GPAs) have positively correlated with higher post-graduation earnings (Jones and Jackson 1990). Of course, given the societal and economic benefits of high grades, college students and their families have developed an interest, and in some cases a mania, with getting higher grades, even if those scores do not accurately reflect real learning. In the past, if students weren't able to earn the highest of grades on their own, they could hire extra-curricular help; perhaps in the form of paid tutoring services or, if they perceive that course content is not worth learning, outsourcing graded work to others via "contract cheating" (Newton 2018). Many contract cheating businesses offer free trials, so even students who are strapped for cash can use those services. Dramatic changes in course delivery, such as the move to remote classes during the Covid-19 pandemic, resulted in a massive increase in use of (and reliance on) third-party "homework helper" sites by students (Lancaster and Cotarlan 2021). Of course, unless the student learns the course content via these outside "helpers" instead of simply copying the answers, these methods sever the link between good grades and student understanding. If extracurricular, outside helpers *do* help students learn, though, fairness dictates that all students have equal access to them, even beyond free trials. After all, the point of education is student learning, whether the locus of learning is a classroom or an online application. Is that where AI comes in?

## AI and Learning

Anecdotally, new iterations of GenAI products are the most recent in a long line of technological advances to disrupt instructors' abilities to assess student learning. Instructors often adapt their courses to account for new technologies. The information-retrieval transition from libraries to online search engines was one such learning era shift (Agosti 2007). Paper and pencil notetaking has largely been replaced by digital notetaking (Mueller and Oppenheimer 2016). Development of online learning management systems has enabled asynchronous and remote class offerings. While each of these technological evolutions have been integrated into college classrooms with the purpose of increasing the convenience and efficiency of student learning, they did not fundamentally change the need for students to pay attention, learn, and rely on their own efforts in order to demonstrate that learning to those who assess their progress.

But GenAI products are novel in that, in addition to providing one-on-one tutoring, they enable students to easily generate content that indicates subject mastery without the previously requisite attention, learning, or effort. Many instructors classify AI use as an academic honesty violation, as it falls within the traditional definition of cheating. But the convenience of information synthesis offered by the products cannot be ignored. Hence, there are myriad debates about the extent to which GenAI products should be allowed in classroom settings.

In the literature, the possibilities inherent in GenAI proliferation are documented with varying degrees of optimism. Humphrey et al have shown that in education as well as in other occupational fields, "AI elicits high-intensity ambivalence, characterized by strong positive and negative emotions coexisting rather than offsetting one another" (2026). Some researchers have taken a full-steam-ahead, all-in view ("generative artificial intelligence [is] a transformative force, revolutionizing education," (Bahroun et al. 2023)), stating that AI makes education "more personalized, engaging, and efficient" (Harry and Sayudin 2023). Some researchers note that the previously discussed need for fairness for disparate student groups to compete with well-off students (who have access to individualized instruction via tutors) is finally met by the availability of LLMs to provide similarly tailored guidance: "AI-powered tutors can bridge educational gaps in underprivileged regions by providing access to quality learning resources" (Ali 2024). GenAI companies themselves are eager to be incorporated into the educational landscape: OpenAI (the maker of ChatGPT) has partnered with several universities (Palmer 2025) and offers discounts to students (OpenAI Help Cent., n.d.).

But not all educators are inclined to invite GenAI into the classroom. The severity with which student use of LLMs can hinder learning and skew the outcomes of traditional assessments has generated substantial opposition to its use in classrooms. Researchers have documented substantial concerns about the technologies inhibiting student learning (Bastani et al. 2024). A recent news article emphasized educators' displeasure with student use of GenAI products; after interviewing a dozen professors, the author concluded that "most scholars see AI as a unique threat, one that extends far beyond cheating on homework and casts doubt on the future of higher education itself in a fast-approaching machine-dominated future" (Speri 2026). Teacher and professor reports of students turning in fully AI-generated content instead of doing work themselves abound in Op-Eds and news articles (Koebler 2025). A study including "over 500 students, teachers, parents, education leaders, and technologists across 50 countries [and] a close review of over 400 studies" prompted researchers to conclude that "at this point in its trajectory, the risks of utilizing generative AI in children's education overshadow its benefits" (Burns et al. 2026). Cognition researchers conducted EEG tests and showed that "LLM users consistently underperformed at neural, linguistic, and behavioral levels" when compared to participants that did not use external tools to complete assignments (Kosmyna et al. 2025). The American Association of University Professors published a report criticizing GenAI products as "untested and unproven" in educational settings, even as they are currently being "adopted uncritically" (Paris et al. 2025). Some professors, after trying to incorporate the technology in critical ways, have been overcome by frustration and quit the profession entirely (Livingstone 2024).

While many educators pool around the poles of GenAI boosterism or criticism, many more are hesitant to pick a side. The novelty of the technology inherently suggests that, while the pros and cons it exercises on student learning are still being established, one shouldn't definitively embrace GenAI or ban it. It is unsurprising, then, that many researchers have taken a "balanced" approach to the use of GenAI products in education, cautioning instructors, students, and institutions to avoid all-or-nothing policies when deciding to adopt or reject the technologies (or asking their students to adopt or reject them).

The endorsement of an anodyne, middle ground response to LLMs, using the GenAI products in conjunction with, or as supplements to, learning, is popular in research literature. Early research into the phenomenon posited that "the use of Generative AI should be supplemented with detailed guidance and flexible strategies" (D.-X. Hu et al. 2025). The suggestion to use GenAI in traditional classrooms is often accompanied by a declaration of the continued necessity of critical thinking skills ("This integration necessitates a dual approach to learning: educating ourselves both about and with GenAI while continuing to develop critical thinking, problem solving, self-regulation and reflective thinking skills" (Yan et al. 2024). Research conducted by administrators (and a marketing professor) at California State University, Fullerton was especially emblematic of the ubiquitous call for balance ("we need to… avoid overzealous use of AI at the risk of impeding human creativity and academic integrity" (Dabirian et al. 2025) while simultaneously insisting upon universal adoption:

> the community would not be served well if … universities do not invest to facilitate access to AI tools… it is absolutely necessary for higher education to incorporate AI… we need… to channel skepticism into at least 'guarded experimentation' of AI to adapt to today's student needs, workforce demand, and technological advancement. (Dabirian et al. 2025)

Dabirian's work arguably highlights a disingenuous ideological stance: everyone must adopt AI use now to avoid doing a disservice to students and the workforce, but shouldn't be overzealous about it. Echoing Dabirian's equivocal AI boosterism is a research article published in the same edition of IEEE Computing Edge: Xu et al. lauded AI use in education to "harmonize human and AI capabilities, fostering a dynamic, inclusive, and adaptive educational ecosystem," insisting that "this balance is crucial" (2024).[1]

While educational ideologies and potential GenAI integration in classrooms is debated and studied by researchers and educators, GenAI products are already available to (and very popular with) students around the world. GenAI companies have no qualms about offering their services and subscriptions to students while experts debate the outcome of the technology on students' brains; likewise, few students are waiting for definitive empirical research before they delegate large swathes of their education to GenAI. A 2024 survey showed that "86% of students use AI in their studies" (Kelly 2024).

## Grade Reporting in the Post-LLM Era

The GenAI-enabled ability for students to indicate mastery of subjects without actually learning them has myriad consequences, but one area of immediate focus has been grading systems. OpEds have decried universities clinging to stuffy, calcified A-F grading systems (the suggestion for doing away with traditional grades was described as "replacing rigor mortis with rigor" (Pink 2025). They posit that universities' insistence that "GPAs are a meaningful metric of achievement" is "misleading" (Mounk 2025). Recent proliferation of LLM products has resulted in renewed frustration in traditional grading systems: "For too long, higher education has focused almost exclusively on outcomes – polished essays, final exams and products that AI can now imitate convincingly, at least on the surface" (X. Hu 2025). To complicate matters, the Covid-19 pandemic (less than three years before the first ChatGPT release) upended traditional lessons and assessments when schools abruptly moved class online, and teachers and professors found themselves accommodating all sorts of academic shortcomings due to the extenuating circumstances inherent in a global pandemic. The end result, as posited by various researchers, was widespread grade inflation (Goldhaber and Goodman Young 2024; Tillinghast et al. 2023; François and De Witte 2025). For these reasons, we included the quantitative component of the study: analyzing grade data over the span of ten years.

---

[1] The authors of these articles advising "balance" in AI in education have more in common than just the middle ground approach. There are indications that the articles themselves were written at least in part by GenAI programs. Suggesting a "balanced approach" is a hallmark of ChatGPT advice (Akhtar 2024). The articles also follow ChatGPT trends in word usage. For example, Leppänen et al. show that ChatGPT output is 10x more likely to use the word "foster" than writing generated by humans (2025). Bahroun uses the word "foster" twelve times in their 2023 paper and Hu et al. (2025) use it four times.

There is clearly a spectrum of reasons for pedagogical prohibition or acceptance of GenAI use in classrooms. However, there is a dearth of research that combines qualitative and quantitative evidence at the student and instructor level for those choices in university settings. So far, most articles documenting the effect of GenAI use on learning are short-term, intervention-based studies (D.-X. Hu et al. 2025), rather than the real-world, long-term, imperfectly measured use that is prevalent in the present day. Some studies documented GenAI usage in particular contexts such as video (Pellas 2025) or lesson planning (Karaman and Göksu 2024), others focus on instruction of AI as a subject rather than its use in classroom settings (Song et al. 2022). Bastani (2024) documented empirical evidence of problematic GenAI use in a single subject (mathematics) in a single high school in Turkey. Kucuk studied the use of GenAI products at a grammar school in Iraq (Kucuk 2024). Some studies document grading patterns in the LLM era but do not include qualitative data from, for example, student and faculty surveys: Researchers in Israel showed that, between the years 2018 and 2024 at a university there, grades changed during the semesters that AI use became widespread: "In 2022-23, average grades in AI-compatible courses rose by one point on a 0-100 scale, growing to 1½ points in 2023-24. The 25$^{th}$ percentile student gained two to three points. Failure rates fell by one-third" (Hausman et al. 2025).

No studies to date have documented the AI learning phenomenon via simultaneous examination of grade data, faculty surveys, and student surveys. There is a gap in the literature regarding changes in post-GenAI exposure instructor level grading, perceptions and assignment design; university-level, multi-subject documentation of student adoption rates and perceptions of GenAI use for learning (or grade trends that may reflect that learning) is also missing.

## Study Design and Methods

In order to answer these research questions, we employed a mixed-methods, multi-course study using retrospective quantitative analysis, instructor surveys, and anonymous student surveys. The retrospective quantitative analysis portion comprised up to ten years of archived grade data across approximately fifteen courses, each taught by a different instructor. Some instructors have been teaching a single course at Umass Boston for over 10 years, so their grade data is fairly comprehensive. Of course, for professors who began teaching at Umass Boston more recently, fewer years of grade data were documented and reported. This information was provided by the university registrar after faculty members gave explicit consent for this portion of the study.

Instructor surveys were designed to glean details and context about professors' decisions to modify their courses in response to students' potential use of LLMs. The student survey included questions about the frequency with, and purpose for, which they used LLMs, how they felt about using LLMs as students, and how they perceive LLM availability has affected their learning experience. Both surveys were designed to measure four disparate concepts in student learning: LLM usage; study behaviors and time use; conceptual understanding and transfer, metacognition, and calibration; and perceptions of grades as a metric to measure learning.

Table 1: Conceptualization of Composite Categories for LLM-Era Perceptions of Student Learning, Studying, and LLM Use

| *Concept* | *Definition* |
|---|---|
| LLM Usage (**LLM**) | The frequency, conditions, and reasons for using LLM-based generative AI products like ChatGPT. |
| Study Behaviors and Time Use (**Study**) | Self-reported study habits: duration and frequency, methods of studying or practicing course content outside of the classroom. |
| Conceptual understanding and transfer, metacognition, calibration (**ConCal**) | Ability to explain and apply course concepts beyond given examples and end-products, awareness of that ability; frequency and method of student's own verification of understanding beyond instructor assessment. |
| Perceptions of grades as a metric to measure learning (**Grades**) | Sentiment regarding the validity and accuracy of grades as a metric by which to measure learning and comprehension. |

Surveys were the best way to answer our research questions because we aimed to compare grading data with faculty and student attitudes of the LLM-available learning environment. Survey research accomplishes this goal "by looking at variation in [a certain] variable across cases, and looking for other characteristics which are systematically linked with it... it aims to draw causal inferences by a careful comparison of the various characteristics of cases" (De Vaus 2013, 5). We employed Likert scale answer choices in order to gather data quickly from large numbers of respondents, effectively documenting participant sentiment without asking those participants for detailed, time-consuming explanations. Another benefit of using Likert scale answer choices is that "the validity of the interpretations made from the data they provide can be established through a variety of means, and the data they provide can be profitably compared, contrasted, and combined with qualitative data-gathering techniques, such as open-ended questions…" (Nemoto and Beglar 2013).

## Content Validity and Interrater Reliability for Survey Instruments

In order to establish "the accuracy with which [our] case study's measures reflect the concepts being studied" (Yin 2014, 238), we conducted an internal content validity survey. We used this survey to ask experts to help us determine the degree of agreement between our survey questions and the concept that the questions designed to measure. Since our study is university education-based, we determined that university professors were the most qualified as experts for the subject matter. We asked five university professors to what extent the survey questions matched the content categories. Interrater reliability was measured by the percentage of agreement among those five raters. Based on their ratings, the ratings for each concept's agreement were as follows: The initial total rating average was 92.93%. The average rating for the "LLM" concept was 96.00%. The average rating for the "Study" concept was 90.00%. The average rating for the Conceptual Understanding and Calibration ("ConCal") was 85.71%. The average rating for the "Grades" concept was 100.00%. Overall, experts were confident in their ratings as measured by the strength ratings, which averaged 2.81/3.

Two items the ConCal category had low agreement (40-60% agreement), initially leading to a lower average rating for that category (85.71% vs 92.93% overall). The original average percentage accuracy for those items was 50% with strong confidence ratings, indicating low validity. As a result, those items were removed from the survey. After deleting those items, the agreement percentage for the ConCal category was 90% and the overall rating average was 94.00%. The content validity table, including rejected items, is on our research website.

## Research Site and Recruitment

### *Research Site Selection*

We deployed this study at the University of Massachusetts, Boston. This university was selected for convenience: both authors work at the university and were able to utilize its communications structures for survey recruitment with few obstacles. Still, the university boasts substantial diversity in both faculty and student populations (UMass Boston, n.d.), which suggests strong potential for generalizability of results. It is hoped that this single-site study may serve as a pilot study for dissemination at regional or even national scale.

### *Recruitment and Selection*

We recruited faculty participants via email from departmental email lists. Research procedures for the surveys took place online via the Qualtrics website. Grade analysis came from course records requested from the registrar with faculty participant consent and exclusion of student personally identifiable information.

Faculty members were sent a preliminary Faculty Participant Recruitment Survey. In that survey, we asked faculty participants to attempt to recruit at least 10 of their students from this past semester for the student survey, so the majority of student outreach was done by faculty participants. The first "question" in each survey (faculty and student) served as the consent form, with faculty participants signing below the consent information in order to proceed to the survey and student participants indicating consent by clicking the “next” arrow (student signatures could have been used to identify them, so they were avoided). The question and display logic used in the faculty and student surveys can be found on our research website.

## Results

### Instructor Survey Results

The primary purpose of this survey was to identify trends and changes in professors’ course structures, grading criteria, and perspectives as they enter the era of student access to GenAI products. Respondents’ answers were analyzed to identify potential relationships between course content, objectives, and grading criteria in different semesters.

Over the course of four weeks in late 2025 and early 2026, recruitment surveys were sent via email to 492 professors at UMass Boston. 57 professors responded to the recruitment survey, and eighteen went on to complete the research survey. The result was a 3.66 % response rate. Considering the survey did not have a paid incentive for participation, the low response rate is unfortunate but understandable. Respondents represented several disciplines, including four Computer Science instructors, two Physics instructors, two Information Technology instructors, two Engineering instructors, and one instructor each from Communication, Women’s and Gender Studies, English, and Biology departments.

### *Key Findings*

- Most instructors modified their courses in response to student use of GenAI.
- Assignment redesign was substantial for many instructors, with the most common estimate being 20–40% of assignments changed.

- Instructors primarily adapted by shifting toward in-class work, in-class assessments, and conceptual reasoning tasks that are harder to complete using AI.
- The severity of assignment redesign increased during the COVID transition and shows a renewed upward trend after Spring 2023, when generative AI tools became widely available.
- Courses taught by the two instructors who reported not changing their course design show a substantially larger increase in average student GPA after the emergence of LLMs (+0.25 GPA points) compared with other instructors (+0.01).

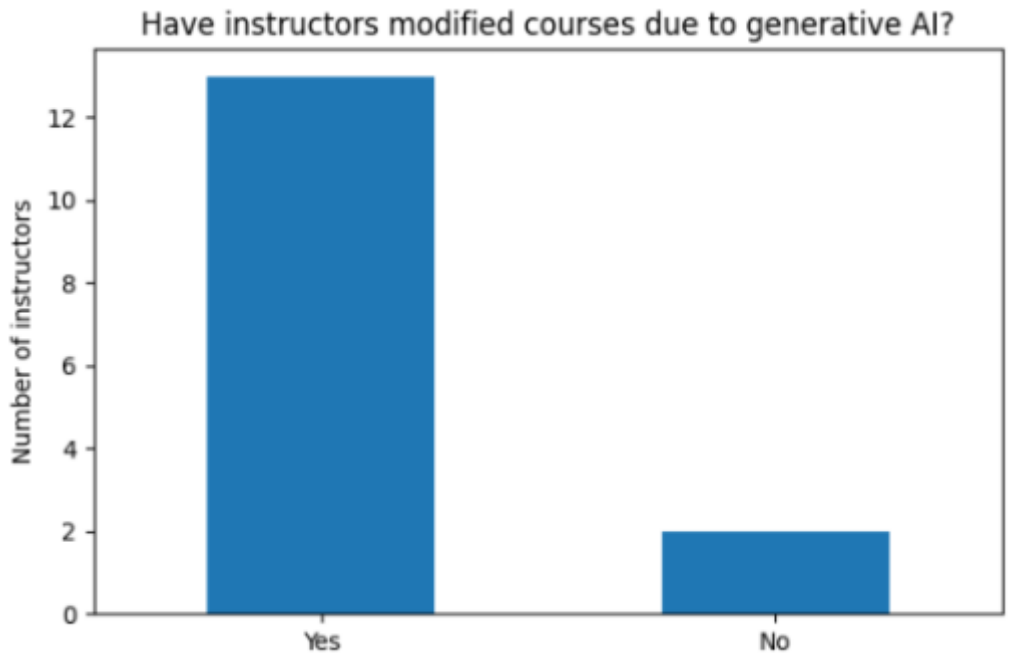

Figure 1: Instructor Modification of Courses Due to GenAI

A large majority of instructors (thirteen of fifteen respondents, 86.7%) reported modifying elements of their course in response to the availability of generative AI tools, while only two instructors reported making no changes.

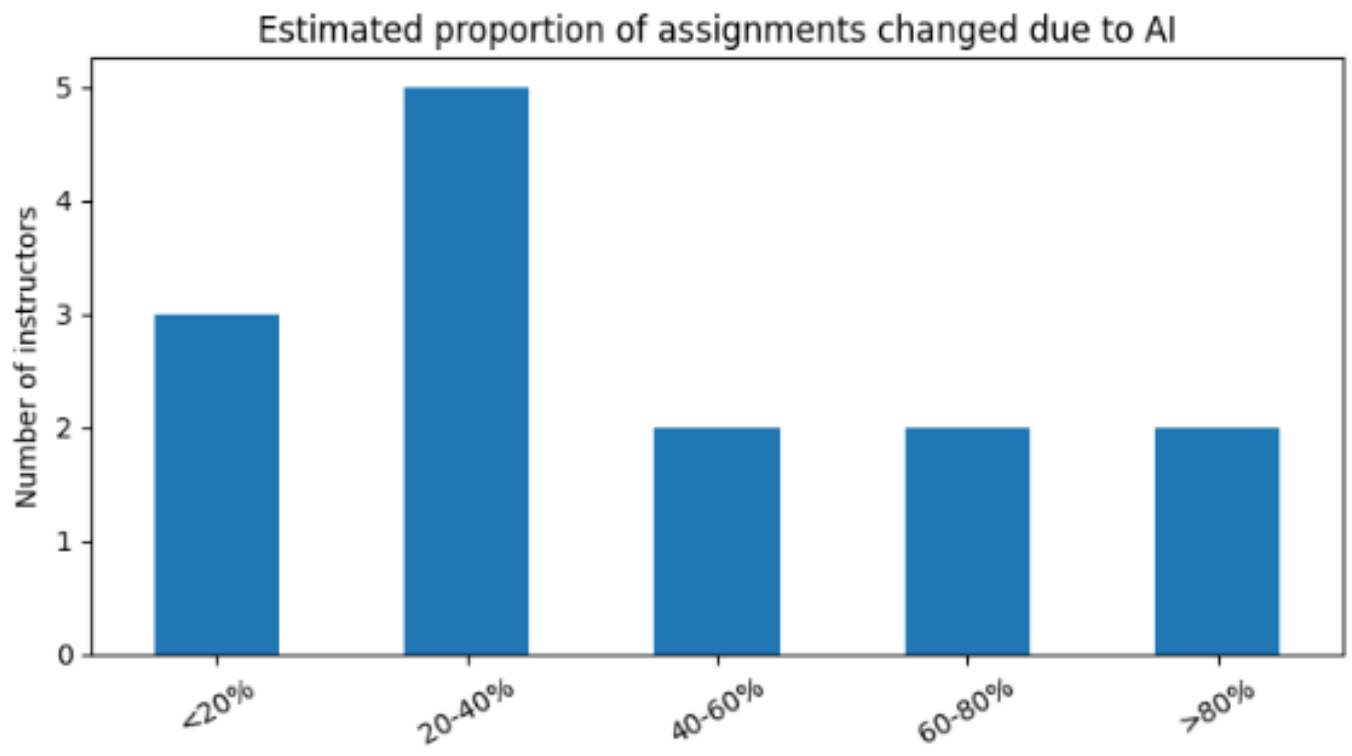

Figure 2: Estimated proportion of assignments changed due to student use of GenAI

Among instructors who reported modifying their courses, the most common estimate was that 20–40% of assignments were changed due to generative AI concerns (five instructors). Several instructors reported even larger adjustments, with some indicating that more than 60% of assignments had been modified.

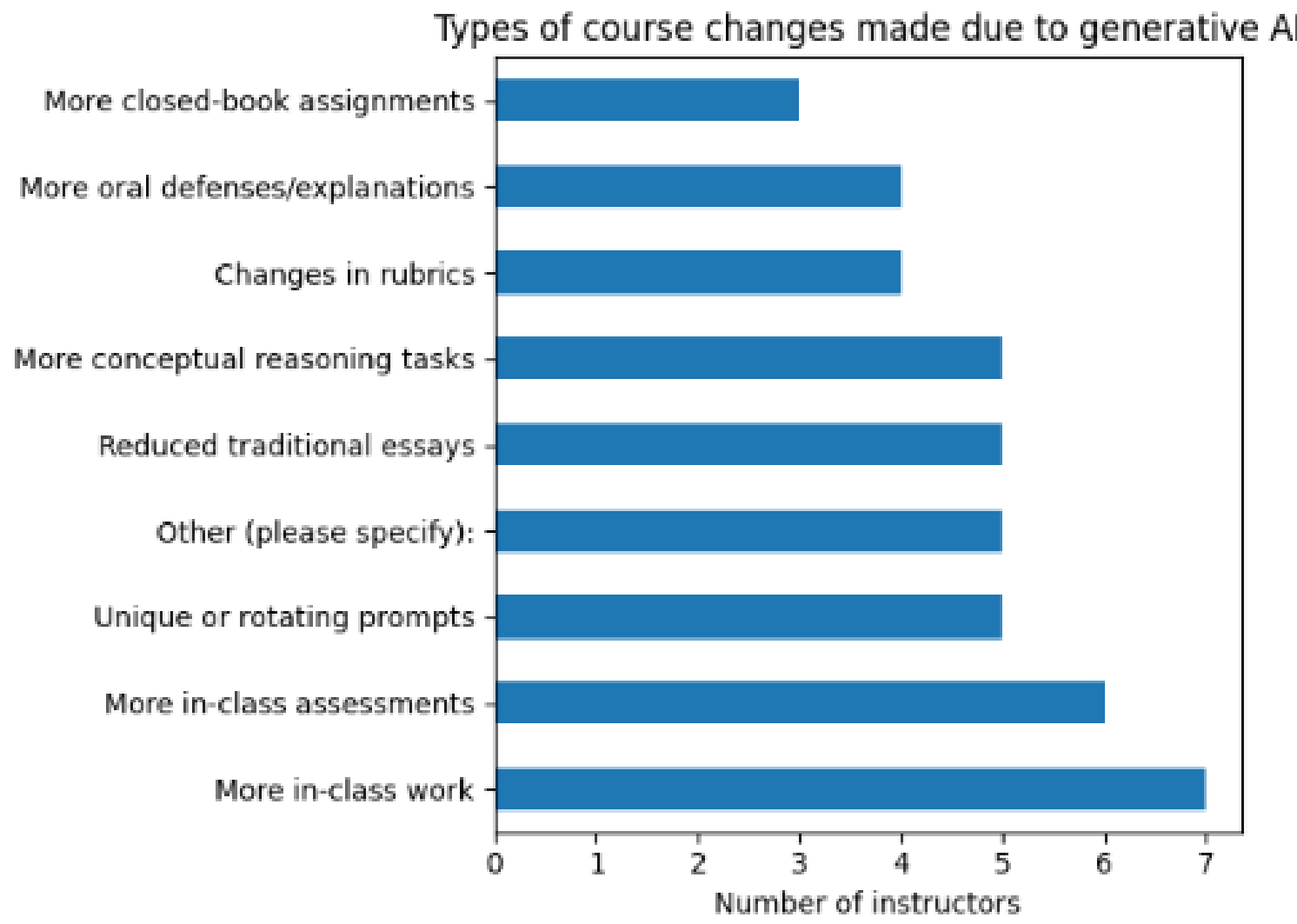

Figure 3: Type of course changes made due to GenAI

Instructors most commonly responded to generative AI by shifting assessments toward activities that are more difficult to complete with AI assistance. The most frequent changes included increasing in-class work (seven instructors) and increasing in-class assessments (six instructors). Several instructors also reported adopting unique or rotating prompts, with emphasis on conceptual reasoning tasks, and reducing traditional essay assignments.

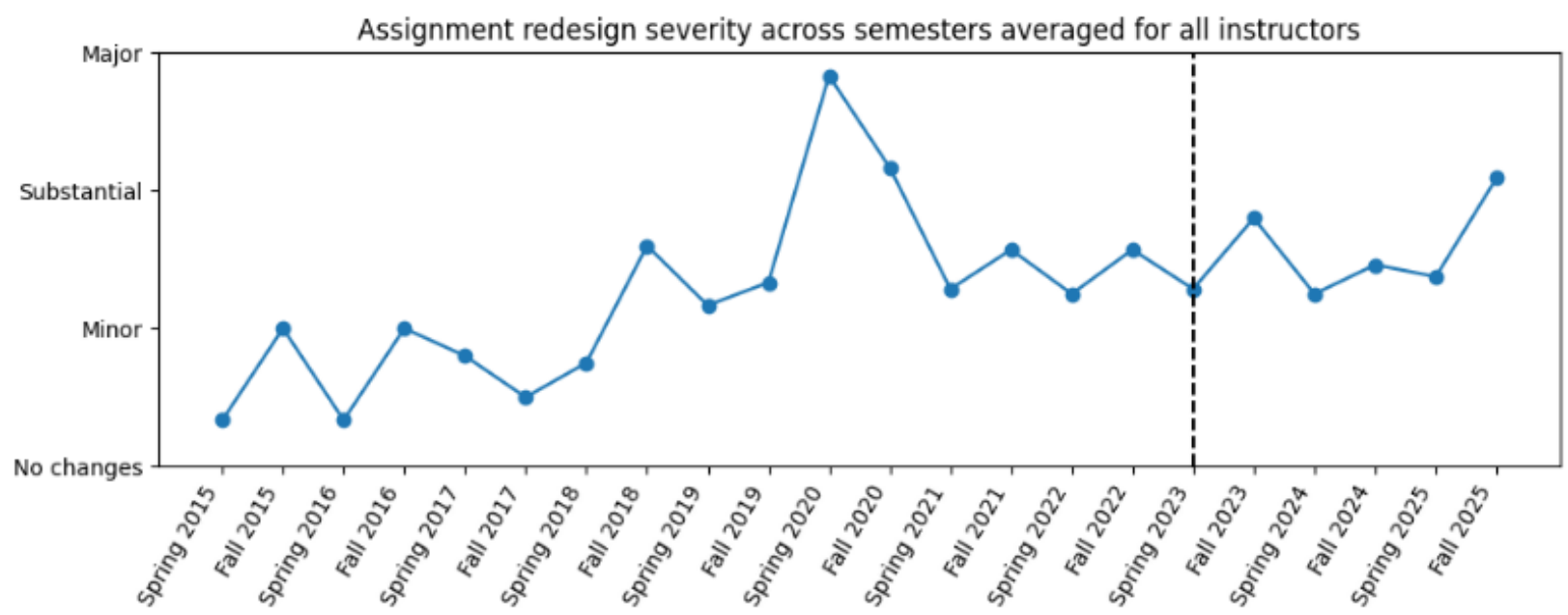

Figure 4: Assignment redesign severity across semesters, averaged for all instructors

Figure 4 shows the average severity of assignment redesign across semesters using a weighted change index (0 = no changes, 1 = minor changes, 2 = substantial changes, 3 = major changes). Assignment changes were generally minor prior to 2020, increased substantially during the COVID-19 transition to remote instruction, and show a renewed upward trend after Spring 2023 during the period when GenAI tools became widely available (indicated with dotted line).

Courses taught by the two instructors who reported not changing their course design show a substantially larger increase in average GPA after the emergence of LLMs (+0.25 GPA points) compared with other instructors (+0.01). Withdrawal rates also declined more strongly in these courses (5.82% → 4.03%) than in courses taught by instructors who redesigned their assignments (4.85% → 4.18%).

### *Thematic Analysis of Faculty Free Response Answers*

In a free response survey item, faculty were asked, “Is there anything else you’d like the researchers to know for this study?” The answers to that question were coded for thematic analysis using Braun & Clarke’s six phase method with the goal of “identifying, analysing and reporting patterns within data” (Braun and Clarke 2006). In this study, we used data-driven, inductive analysis, meaning the themes were derived from data rather than aligning the data with pre-defined themes. “Inductive analysis is … a process of coding the data without trying to fit it into a preexisting coding frame, or the researcher’s analytic preconceptions… this form of thematic analysis is data-driven.” The following themes emerged for faculty free response answers:

**Flux:** trial and error. Answers indicating that faculty were currently involved in or had recently finished making changes to their courses due to student AI use were included in this category. Emblematic quotes for this category include the responses “The likelihood of students using AI in my class is reduced because half of my 'homework' was always done in-class due to students cheating before AI. Due to AI, I am going to remove the remaining half of homework…” (Physics professor), “I still need to change more assessment to be process and thinking focused and less on the final product. It's going to take me a while, but I've started” (Engineering professor), and “I tried for a couple semesters to convince them not to use AI… eventually I gave up and switched to … in-class assessments” (Computer Science professor).

**Variety in student adoption and admission**. Answers indicating that faculty were experiencing a wide range of student outcomes as they pertain to AI use in class were included in this category. Representative quotes for this category include the responses “They were required to give a log of when they used AI but sometimes they would not admit it, other times they just gave a partial log and other times they gave the full log” (Biology professor), and “I got a sense of how each student decided to deal with the opportunity to use chatgpt in our/their work. Some ignored it, some did well, some did medium" (Computer Science professor).

**Working with AI**. Answers indicating that faculty were voluntarily adapting to and even embracing student AI use in class were included in this category. Emblematic quotes for this category include the responses "I think we need to embrace AI not fight it" (Computer Science professor) and "[GenAI is] not banned and, in many cases, we're exploring to what degrees the tools do and do not help them reach their goals… In my class, we create an AI policy together based on what the students want" (English professor).

**Frustration**. Interestingly, only one professor out of eighteen used the free response question to voice frustration with student use of GenAI. They complained of AI use forcing assignment shifts from high creativity but less verifiability to the opposite:

> My course used to be known as one that made students think… with creative and lively discussion of articles [they] read at home. In November, 2022 I started getting the AI slop answers, even to personal questions. I tried for a couple semesters to convince them not to use AI, but I was not successful... eventually I gave up and switched to more boring, straightforward in-class assessments… (Computer Science professor)

## Student Survey Results

Student respondents were primarily drawn from computing-related programs, including thirteen students from Computer Science or Information Technology and six students from Engineering.

### *Key Findings*

Analysis of student survey responses yielded the following key findings:

- Students underreport their own LLM use on graded assignments.
- Reported LLM use spans both graded assignment and learning contexts.
- Higher LLM users are less likely to report that grades accurately reflect learning.
- Reported LLM use does not correlate with GPA, study time, or conceptual understanding.

As expected, student responses about the use of LLMs indicated social desirability bias. Students reported substantially lower personal LLM use for graded assignments than the level of LLM use they attributed to their peers, suggesting systematic underreporting in self-reported AI use for graded assignments.

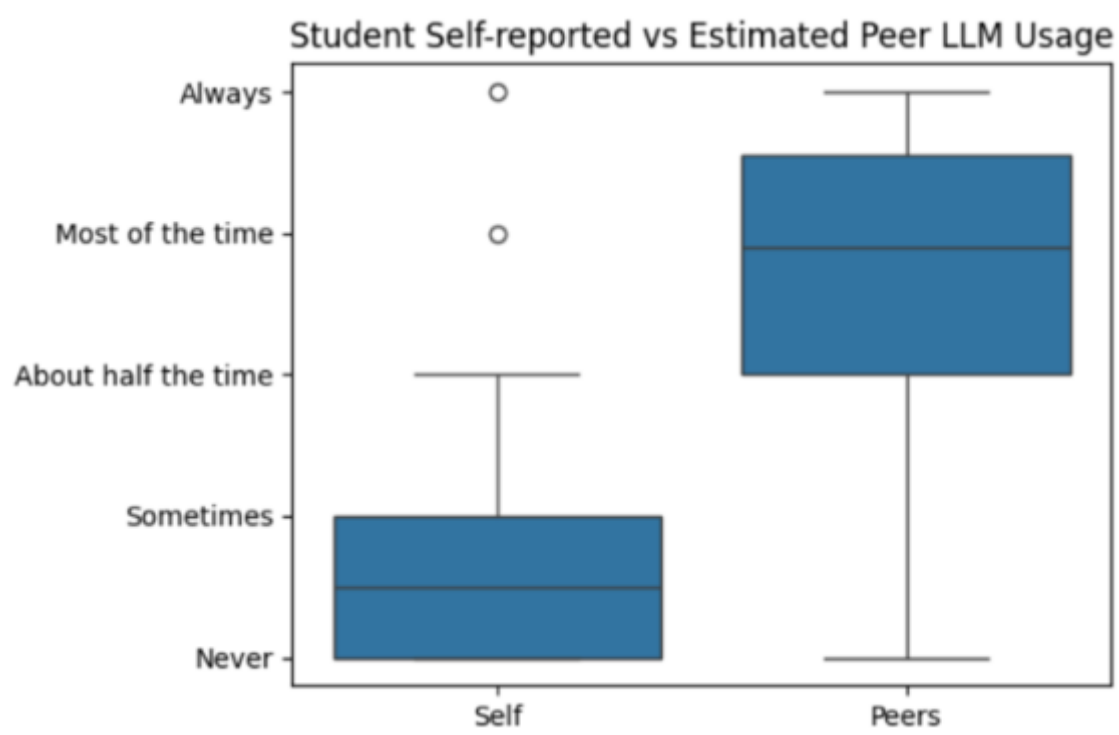

Figure 5: Student LLM-usage underreporting index

The remainder of student response analysis plots survey responses in relation to the conceptualization framework for composite categories established in Table 1.

Students who reported higher LLM use for graded assignments also reported higher LLM use for learning concepts (Spearman r = 0.52, p = 0.033), indicating that LLM use tends to occur across multiple learning contexts (e.g. studying) rather than "just" assignment completion. Higher LLM use was also associated with a lower likelihood of reporting that grades accurately reflect learning (Spearman r = –0.52, p = 0.031): students who rely more on LLMs view grades as a less reliable indicator of their actual understanding.

For study behaviors and time use, higher self-reported LLM use was weakly associated with reduced practice without AI (Spearman r = –0.32, p = 0.19), which suggests that there is partial substitution between AI-assisted work and independent practice, though the relationship was not statistically significant. LLM use for graded assignments was not associated with the number of hours students reported studying (Spearman r = 0.08, p = 0.76), indicating that LLM use does not appear to meaningfully alter overall study time.

We did not observe a relationship between self-reported LLM use and the ConCal composite measuring conceptual understanding, transfer, confidence, and self-checking (Spearman r = –0.16, p = 0.52). Nor was students' self-reported LLM use for graded assignments associated with students' GPA (Spearman r = 0.03, p = 0.91, n = 15), indicating no relationship between LLM usage frequency and academic performance in this sample. The ConCal composite did show a moderate positive relationship with GPA (Spearman r = 0.44, p = 0.10), suggesting that the ConCal composite reflects academic performance trends, though the relationship was not statistically significant in this sample. 25% of respondents said they have changed their study habits to rely on GenAI products in the post-LLM era.

All student respondents had at least some concerns about GenAI in the post-LLM era. Many were concerned about being accused of GenAI use when they didn't use it; fewer were worried about being "caught" when using it.

Table 2: Student responses to the question "Which, if any, concerns do you have about attending university classes in the age of generative AI? Please select all that apply"

| | |
|---|---|
| I'm concerned that my professors will punish me for using AI when I use it. | 33% of respondents |
| I'm concerned that my professors will punish me for using AI even when I don't use it. | 61% of respondents |
| I'm concerned that I won't learn as much as I would if AI weren't available. | 39% of respondents |
| I'm concerned that jobs I would qualify for after graduating will be replaced by AI. | 61% of respondents |

Holistically, the student survey responses suggest that while LLM use is common and occurs across multiple learning contexts, it does not appear to be strongly associated with differences in study time, conceptual understanding, or academic performance as perceived by students in this sample. This result appears to contradict data from the faculty survey, which is not completely surprising: the views of professors with pedagogical expertise about student use of LLMs differ from the perceptions of students who are still cultivating new learning skills. Student concerns about GenAI do not appear to mitigate their use of it.

### *Thematic Analysis of Student Free Response Answers*

Using the same thematic analysis method as we did for faculty free response items, student free response answers were thematically coded. Emerging themes around student LLM use include staying competitive, critical views, and the need for institutional guidance.

**Staying competitive** (market and intellectual). Answers indicating that students were focused on being competitive in the post-graduation job market and seeking better grades were included in this category. For example, one student noted "Since ai is the future, the students should not be completely ignoring it if they want to stay competitive." One student also voiced concern that not using AI was jeopardizing their learning:

> Receiving lower grades while not using AI can be discouraging when I overhear someone use AI for their better grade. But talking out the solutions with a student afterwards and seeing how much they learned because they used AI to fill in the blanks makes me question if I did the right thing by not using it. (Student free response)

**Critical views**. Some students voiced displeasure with AI in their educational pursuits; those answers were gathered in this category. Representative quotes include "I think AI is a good tool for the efficiency of the industry, but it cannot improve the human intelligence in education" and "I dont think AI is a good influence since it can provide wrong answer and you wont know what's right and wrong."

**Need for institutional guidance**. Several students indicated that the university should provide more structured guidance on AI use. Those answers were gathered in this category. Emblematic quotes include “It makes me wonder how different it would be if the school provided an A.I. model/source that references approved material but doesn't just give answers away either,” and “There should be a course that teaches you how to use ai and how not to use it, that stresses the importance of actually learning the topics your future courses cover rather than just using ai to do the assignments**.**” Another student contextualized the need for this type of guidance with examples of other critical university interventions:

> I think [students] should be made aware of what is the correct and incorrect use of it as a student through either mandatory workshops like the ones we have for phishing and cyber attacks or a topic to be taught in a course that practically makes them implement and use ai in a good vs bad way. (Student free response answer)

## Grade Data Results

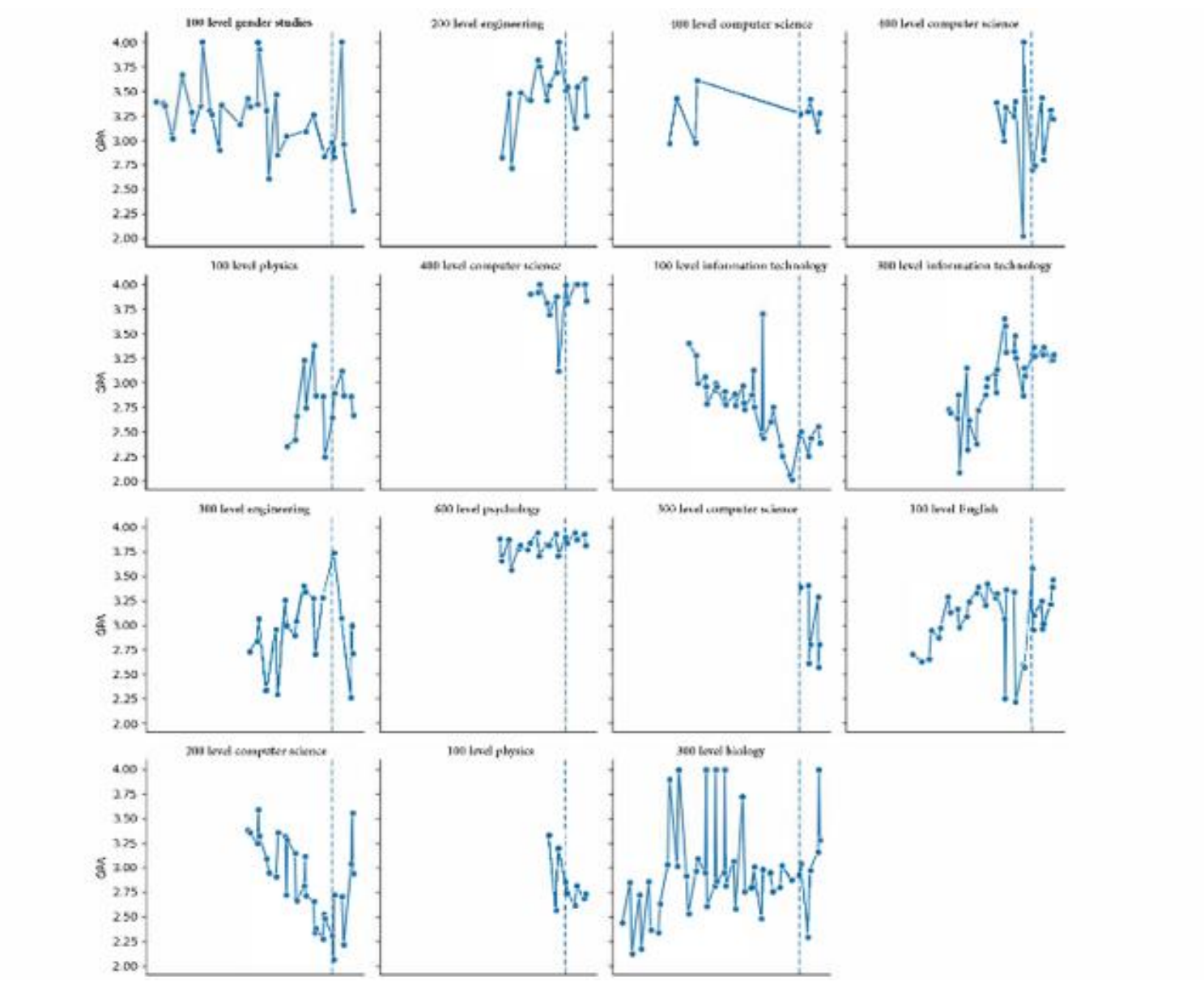

Figure 6: Individual instructor grade trends over time

Spring 2023, the first full semester in which students had access to GenAI products is classified in this study as the first semester in the Post-LLM period and is indicated with a dotted vertical line in the above graphs. Comparing the Pre- and Post-LLM periods through the lens of student grades reported to the registrar by participating faculty, the following trends are present: Student grade point averages are slightly higher, withdrawals are slightly lower, and failures are slightly lower.

Mean student GPA in individual courses increased slightly from 2.96 (pre-LLM) to 3.00 (post-LLM). Although this difference was statistically significant (Welch t-test, $p = 0.0078$), the effect size was negligible (Cohen's $d = 0.04$), suggesting that the practical magnitude of the change is minimal. Withdrawal rates decreased slightly from 4.95% in the pre-LLM period to 4.17% post-LLM. Although the difference was statistically significant (Welch t-test, $p = 0.00055$), the effect size was negligible (Cohen's $d = -0.037$). Failure rates decreased from 3.96% in the pre-LLM period to 2.57% post-LLM. Although the difference was statistically significant (Welch t-test, $p < 0.001$), the effect size was small (Cohen's $d = -0.08$).

## Discussion

Tens of millions of students worldwide are using LLM products as part of their university studies (Kelly 2024). This study, however, included only one university and cannot provide a holistic look at all the faculty and students affected by widespread LLM availability on campuses. The inclusion of only one university limited the results of the current study but also accentuates its potential use as a pilot study for larger populations.

We relied on convenience sampling to select faculty and students for this study, which presents a second limitation. We sent the survey to all faculty at UMB via department chair email lists, and all those who passed the recruitment requirements were included. Student participants were recruited via participating faculty, so students who may have interesting input on the subject matter may have been unintentionally excluded simply because none of their professors participated in the faculty portion of the survey. While the sample size in this study was too small for highly generalizable results, it still yielded some interesting initial insights as faculty and students adjust to the ubiquity of GenAI products in academic settings.

### Faculty Survey Results Discussion

Contrary to suggestions from the literature review, the majority of participating faculty at UMB are not opposed to student AI usage. Faculty changes to course design have increased in the post-LLM era (Spring 2023 onward), with a delay of approximately two semesters. Data suggests that, on average, faculty made relatively few changes to their courses in the first year after widespread LLM availability but dramatically changed their courses to address student AI use starting in Fall 2024 (Figure 4). Most faculty members have made substantial or major changes to their courses, altering over 20% of their traditional course design, as a direct consequence of AI proliferation, with some altering over 80% of their courses. The extra time and resources spent on these changes is difficult to quantify; faculty dedicating time and resources to drastic course design changes may have less time to meet with students for office hours, for example.

The few professors who have not made changes to their courses in the post-LLM era have documented student course averages that are higher than their peers who did change course design in the same timeframe. Combined with the literature-informed and underreporting-index-supported acknowledgement of very widespread GenAI usage, this small empirical finding further fuels the larger-scale debate about student LLM usage discussed in the introduction and literature review. The combination of the literature review and our own survey results suggest that those scores are artificially inflated due to students using GenAI to complete their assignments and meet course objectives rather than doing so themselves. Alternative reasons for this gap are impossible to rule out; the small sample size means that minor shifts in faculty habits or student groups could result in random correlations. Taken at face value, this outcome lends credence to student concerns about falling behind if they don't use LLMs to complete classwork. Indeed, it appears that when faculty do not change their course to be less AI-solvable, student use of AI does result in higher scores and puts non-users at a competitive disadvantage.

None of the instructors who elected to complete the survey taught remote writing courses. These courses appear to be the most challenging to redesign in the LLM era, since the typical adjustments made by faculty in response to student GenAI use were "more emphasis on in-class work" and "fewer traditional essay assignments" - neither of which would appear to be feasible for instructors of online writing courses. Input from these instructors will be important to ascertain post-LLM era pedagogical shifts in the teaching of writing in remote contexts.

## Student Survey Results Discussion

78% of student respondents say they use GenAI products to ask for explanations of concepts. While none of the respondents said they use GenAI products to generate answers for graded work, the underreporting index for student LLM usage in that category is very high. Students are hesitant to admit using LLMs even in an anonymous survey, which supports faculty perceptions of students being dishonest about using them in class. For example, a 200-level Computer Science professor noted, "It's ... concerning that some [students] lie about AI use."

Students are concerned about being competitive in courses and in the workforce and are explicitly requesting, unprompted, university-level guidance on AI in the form of workshops and classes. UMass Boston seems to focus its guidance on AI to a faculty audience. There are two web pages (https://www.umb.edu/learning-design/artificial-intelligence-ai-guidance/ and https://www.umb.edu/academics/provost/academic-integrity/faculty-guidance/#:~:text=Academic%20integrity%20and%20AI) that offer guidance on AI implementation from a faculty perspective, but none that uniquely address student AI use from the student perspective. While the university encourages faculty to allow varying degrees of student GenAI use, the words "LLM" "AI" or "artificial intelligence" are not mentioned in the student code of conduct, which was last updated in August of 2025. As such, all AI guidance appears to be at the individual faculty level. Given students' stated need for guidance on the subject, a GenAI policy and/or workshop for students is recommended at the university level.

### Grading Data Results Discussion

Failure and withdrawal rates are negatively correlated to LLM access, and GPA is weakly positively correlated to LLM access. Using the traditional interpretation of grading schemas, access to LLMs is increasing student performance, or at least the appearance of it. Previous studies indicate that the increase in student performance indicative in this dataset may be attributed to LLMs replacing student cognition rather than improving it (Kosmyna et al. 2025; Bastani et al. 2024). As noted in the faculty survey discussion above, the few professors who have not made any changes to their courses in the post-LLM era have documented student course averages that are 0.24 GPA points higher than their peers who did change course design in response to student LLM access in the same timeframe. At UMB, the difference between logged letter grades (e.g., between C and C+, between A- and A) is 0.30 GPA points. Students in classes with professors who are not adapting their courses in the post-LLM era are enjoying a substantial grade boost.

## Conclusion:

This was the first study of its kind to examine the effects of student LLM usage by triangulating data from faculty surveys, student surveys, and historical grade reports. Despite substantial limitations, the research indicates several patterns within the phenomenon of student usage of GenAI products, including their effects on faculty course design and grade reporting. Most faculty members have made substantial or major changes to their courses, altering over 20% of their traditional course design as a direct consequence of AI proliferation, with some altering over 80% of their courses. For most professors, those changes were necessitated within one year of the release of publicly available LLMs; others have established plans for further adjustments in the coming semester. Professors who have not changed their courses in the post-LLM era report substantially higher student grades. Professors are generally not against student use of GenAI in their courses, with some vociferous exceptions. Students' main concern appears to be GenAI's potential downsides (when they use it) and unfairness (when their peers use it to get better grades and they abstain), and its effect on the post-graduation job market.

## Acknowledgment

This study was supported by the AI Research Core at UMass Boston and Alfred P. Sloan Exemplary Pathways grant G-2025-25149

## AI Acknowledgment

The authors declare that generative AI or AI-assisted technologies were not used in any way to prepare, write, or complete this manuscript. The authors confirm that they are the sole authors of this article and take full responsibility for the content therein, as outlined in COPE recommendations.

## Informed Consent

This study was conducted with the informed consent of all participants. Participants were informed of the study's purpose, procedures, potential risks and benefits, and their right to withdraw at any time without penalty. Written consent was obtained from all participants. The study protocol was approved by the Institutional Review Board at Umass Boston.

## Conflict of Interest

The authors declare that there is no conflict of interest.

## ABOUT THE AUTHORS

**Amanda Potasznik**: Senior Lecturer, Computer Science Department, University of Massachusetts, Boston, Massachusetts, USA
Corresponding Author's Email: Amanda.Potasznik@umb.edu

**Daniel Haehn**: Associate Professor, Computer Science Department, University of Massachusetts, Boston, Massachusetts, USA
Email: Daniel.Haehn@umb.edu